\documentclass[10pt,aps,prl,twocolumn,superscriptaddress,floatfix,showpacs,longbibliography]{revtex4-1}

\usepackage{graphicx}
\usepackage{amsfonts}
\usepackage{amssymb}
\usepackage{amsmath}
\usepackage{txfonts}
\usepackage{lipsum}
\usepackage{color}
\usepackage{wasysym}
\usepackage[colorlinks=true, allcolors=blue]{hyperref}
\usepackage{bbold}
\usepackage{etoolbox} 
\usepackage[normalem]{ulem}
\usepackage{mathtools} 
\usepackage{multirow} 
\usepackage{hhline}
\usepackage{mathrsfs} 
\usepackage{soul}

\definecolor{green}{rgb}{0,0.6,0.1}

\usepackage{letltxmacro}
\LetLtxMacro{\oldsqrt}{\sqrt}
\renewcommand{\sqrt}[2][\mkern8mu]{\mkern-6mu\mathop{}\oldsqrt[#1]{#2}}

\usepackage{url}
\definecolor{indigo(dye)}{rgb}{0.0, 0.25, 0.42}
\usepackage{placeins}

\begin{document}

\title{Coexisting charge density wave and ferromagnetic instabilities in monolayer InSe}

\author{E. A. Stepanov}
\email{evgeny.stepanov@polytechnique.edu}
\affiliation{CPHT, CNRS, Ecole Polytechnique, Institut Polytechnique de Paris, F-91128 Palaiseau, France}
\affiliation{\mbox{Department of Theoretical Physics and Applied Mathematics, Ural Federal University, 620002 Ekaterinburg, Russia}}

\author{V. Harkov}
\affiliation{I. Institute of Theoretical Physics, University of Hamburg, Jungiusstrasse 9, 20355 Hamburg, Germany}
\affiliation{European X-Ray Free-Electron Laser Facility, Holzkoppel 4, 22869 Schenefeld, Germany}

\author{M. R\"osner}
\affiliation{Radboud University, Institute for Molecules and Materials, 6525AJ Nijmegen, The Netherlands}

\author{A. I. Lichtenstein}
\affiliation{I. Institute of Theoretical Physics, University of Hamburg, Jungiusstrasse 9, 20355 Hamburg, Germany}
\affiliation{The Hamburg Centre for Ultrafast Imaging, Luruper Chaussee 149, 22761 Hamburg, Germany}
\affiliation{European X-Ray Free-Electron Laser Facility, Holzkoppel 4, 22869 Schenefeld, Germany}
\affiliation{\mbox{Department of Theoretical Physics and Applied Mathematics, Ural Federal University, 620002 Ekaterinburg, Russia}}

\author{M. I. Katsnelson}
\affiliation{Radboud University, Institute for Molecules and Materials, 6525AJ Nijmegen, The Netherlands}
\affiliation{\mbox{Department of Theoretical Physics and Applied Mathematics, Ural Federal University, 620002 Ekaterinburg, Russia}}

\author{A. N. Rudenko}
\email{a.rudenko@science.ru.nl}
\affiliation{Radboud University, Institute for Molecules and Materials, 6525AJ Nijmegen, The Netherlands}
\affiliation{\mbox{Department of Theoretical Physics and Applied Mathematics, Ural Federal University, 620002 Ekaterinburg, Russia}}

\begin{abstract}
Recently fabricated InSe monolayers exhibit remarkable characteristics that indicate the potential of this material to host a number of many-body phenomena.
Here, we consistently describe collective electronic effects in hole-doped InSe monolayers using advanced many-body techniques.  
To this end, we derive a realistic electronic-structure model from first principles that takes into account the most important characteristics of this material, including a flat band with prominent van Hove singularities
in the electronic spectrum, strong electron-phonon coupling, and weakly-screened long-ranged Coulomb interactions.
We calculate the temperature-dependent phase diagram as a function of band filling and observe that this system is in a regime with coexisting charge density wave and ferromagnetic instabilities that are driven by strong electronic Coulomb correlations. This regime can be achieved at realistic doping levels and high enough temperatures, and can be verified experimentally.
We find that the electron-phonon interaction does not play a crucial role in these effects, effectively suppressing the local Coulomb interaction without changing the qualitative physical picture.
\end{abstract}

\maketitle

Two-dimensional (2D) group III-VI metal chalcogenides have recently attracted great interest because of their appealing  characteristics. 
Among them are high charge carrier mobility, controllable energy gaps, excellent thermoelectric and optical properties, as well as excellent stability at ambient conditions~\cite{bandurin2017high, doi:10.1021/acs.nanolett.5b00493, Kumar2014, Mudd2016, Hung2017, Wei2018}. 
The electronic structure of ultrathin InSe features flat regions in the valence band dispersion leading to prominent van Hove singularities (vHS) in the hole density of states (DOS)~\cite{PhysRevB.89.205416}. 
Importantly, this kind of electronic structure is only observed in the monolayer limit of these materials, as has been experimentally demonstrated by means of angular resolved photoemission spectroscopy~\cite{doi:10.1063/1.5027023, PhysRevMaterials.3.034004}. 
If the vHS appears at the Fermi energy, it may result in numerous competing channels of instabilities such as magnetic, charge, or superconducting order with a very non-trivial interplay between them~\cite{PhysRevB.64.165107, PhysRevB.67.125104, RevModPhys.84.299}.
Scientific interest to flat-band materials has been triggered in recent years by the discovery of unconventional superconductivity and related exotic phenomena in twisted bilayer graphene~\cite{cao2018unconventional, cao2018correlated, Yankowitz1059}.
Therefore, thin films group III-VI materials turn out to be prospective candidates for studying many-body correlation effects, which are closely related to the above-mentioned features in the electronic spectrum~\cite{PhysRevLett.112.070403}, circumventing the need for any twist engineering. 

The theoretical description of many-body effects in the monolayer InSe is challenging and is limited to a few works mainly focusing on the electron-phonon coupling. 
In particular, it has been shown that hole states in this material undergo significant renormalization due to the interaction with acoustic phonons, which gives rise to the appearance of unusual temperature-dependent optical excitations~\cite{PhysRevLett.123.176401}.
Also, a strong electron-phonon interaction may result in a charge density wave (CDW) instability predicted recently~\cite{PhysRevB.103.035411}.
At the same time, monolayer InSe 
is expected to possess strong electronic Coulomb correlations that have not been accurately studied yet.
For instance, the presence of the Mexican-hat-like band in this material might favors a magnetic instability that could lead to the formation of a magnetically ordered state at low temperatures. 
Up to now the existence of a magnetic solution for InSe monolayer has been demonstrated only at the level of density functional theory (DFT)~\cite{doi:10.1021/acsanm.8b01476}.
In addition, the weakly screened Coulomb interaction in 2D may result in a Coulomb-driven CDW instability.
However, a systematic many-body consideration of these effects is still missing in the literature. 

In this Letter, we address the problem of collective electronic effects in monolayer InSe.
For this purpose, we derive a realistic model that considers both, long-range Coulomb interactions and the electron-phonon coupling.
The introduced interacting electronic problem is further solved using an advanced many-body approach that explicitly takes into account non-local collective electronic fluctuations.
In the regime of hole-doping, we find that charge ordering represents the main instability in the monolayer InSe. 
It is formed in a broad range of doping levels and corresponds to a commensurate CDW, which indicates that this instability is rather driven by strong electronic Coulomb correlations than by an electron-phonon mechanism as discussed previously.
Inside the CDW phase, we detect another collective effect that drives the system towards a ferromagnetic (FM) ordering.
This instability is formed in close proximity to the vHS in the DOS.
Finally, we observe that the electron-phonon coupling tends to suppress the FM ordering, enlarging the CDW phase.
However, the presence of the electron-phonon coupling does not qualitatively affect the observed effects. 

\begin{figure}[t!]
\includegraphics[width=1.0\linewidth]{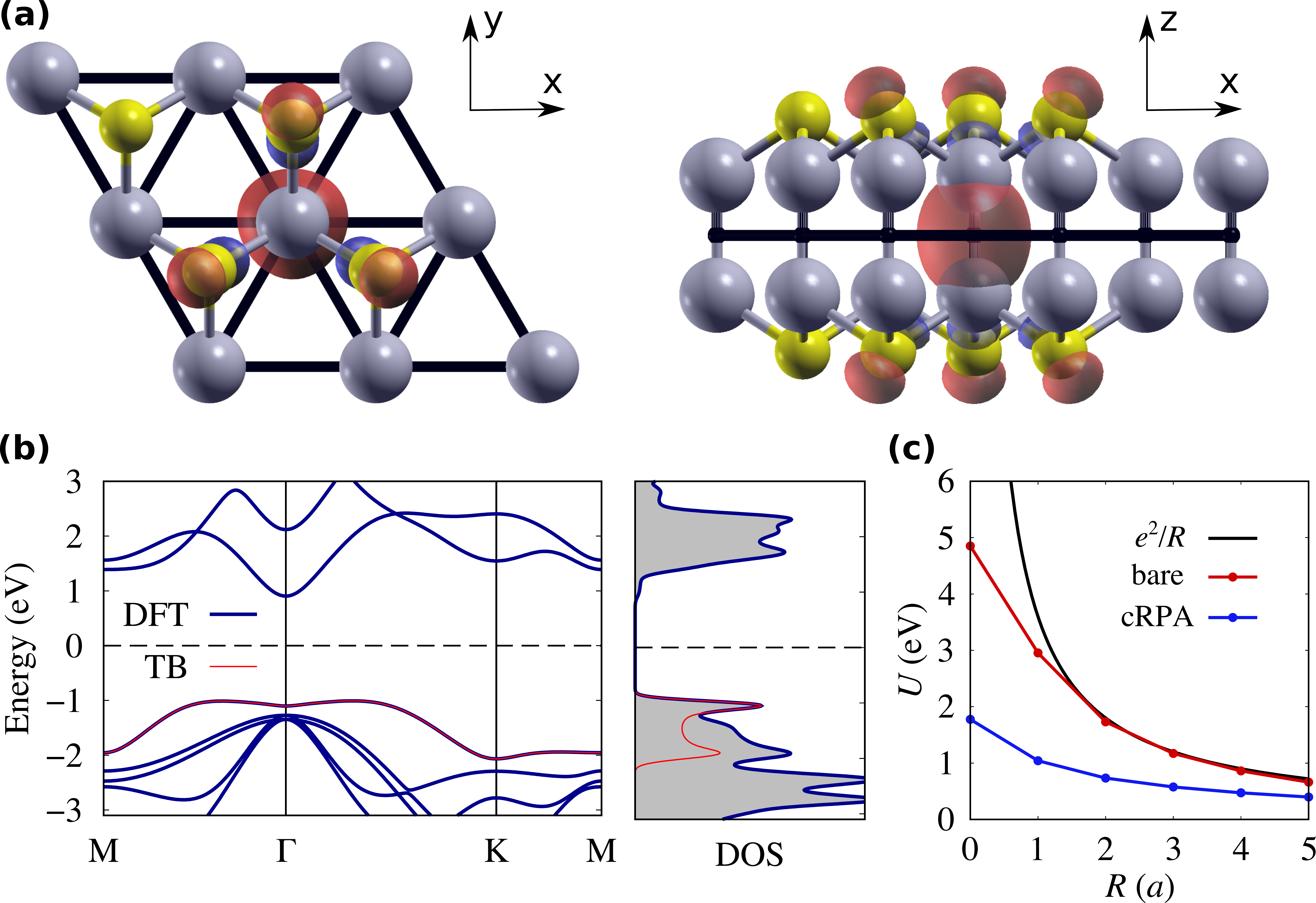}
\caption{(a) Schematic crystal structure of monolayer InSe shown in two projections. Superimposed is an isosurface of the Wannier function describing the valence states in InSe. The black lines depict an effective electronic lattice; (b) Band structure and DOS calculated in DFT (blue), and from tight-binding model parametrization (red); (c) The Coulomb interaction between the Wannier orbitals shown as a function of the distance calculated within the cRPA scheme. The bare (unscreened) Coulomb interaction is shown for comparison. \label{fig:bands}}
\end{figure}

{\it Model} --- 
InSe is a layered van der Waals material, where each layer consists of two vertically displaced In-Se honeycombs, giving rise to four Se-In-In-Se atomic planes [Fig.~\ref{fig:bands}(a)]. In the monolayer limit DFT calculations predict InSe to be a semiconductor with an indirect energy gap of ${\sim2\,\text{eV}}$.
The electronic dispersion shows a single well-separated valence band, which has the shape of a Mexican hat, as depicted in Fig.~\ref{fig:bands}(b). 
This peculiarity is of great advantage for many-body considerations as it allows us to reduce the correlated subspace to a single band. 
To this end, we construct a tractable tight-binding model that accurately reproduces this highest valence band [Fig.~\ref{fig:bands}(b)]. 
The corresponding model is defined in terms of maximally-localized Wannier functions on an effective triangular lattice, as shown in Fig.~\ref{fig:bands}(a). 
Each Wannier function is reminiscent of an In-In bonding orbital with some tails on Se atoms. 
The resulting single-band model Hamiltonian
on a triangular lattice 
reads 
\begin{gather}
H = \sum_{ij,\sigma} t^{\phantom{\dagger}}_{ij} c^{\dagger}_{i\sigma} c^{\phantom{\dagger}}_{j\sigma} + U\sum_{i}n^{\phantom{\dagger}}_{i\uparrow} n^{\phantom{\dagger}}_{i\downarrow} + \frac12 \sum_{i\neq{}j,\sigma\sigma'} V^{\phantom{\dagger}}_{ij}\, n^{\phantom{\dagger}}_{i\sigma} n^{\phantom{\dagger}}_{j\sigma'} \notag\\
+~\omega_{\rm ph}\sum_{i}b^{\dagger}_{i}b^{\phantom{\dagger}}_{i} + g\sum_{i,\sigma}n^{\phantom{\dagger}}_{i\sigma}\left(b^{\phantom{\dagger}}_{i}+b^{\dagger}_{i}\right)
\notag
\end{gather}
where $c^{(\dagger)}_{i\sigma}$ operator annihilates (creates) an electron on the site $i$ with the spin projection $\sigma=\{\uparrow,\downarrow\}$. 
The {\it ab initio} electronic dispersion is reproduced by five neighboring hopping amplitudes $t_{ij}$:
${t_{01} = 127.9\,\text{meV}}$, ${t_{02} = -\,41.8\,\text{meV}}$, ${t_{03} = -\,45.0\,\text{meV}}$, ${t_{04} = 13.0\,\text{meV}}$, and ${t_{05} = -\,4.4\,\text{meV}}$.
Figure~\ref{fig:bands}(c) shows the bare and screened Coulomb interaction between the Wannier orbitals calculated as a function of the distance within the constrained random phase approximation (cRPA). 
The on-site screened Coulomb repulsion ${U=1.78\,\text{eV}}$ greatly exceeds the bandwidth ${\approx1\,\text{eV}}$, which usually indicates well-developed magnetic fluctuations in the system.
As expected for a 2D material, the non-local Coulomb interaction $V_{ij}$ in monolayer InSe is weakly-screened and long-ranged [see blue line in Fig.~\ref{fig:bands}(c)].
Moreover, the interaction $V_{01}=1.04\,\text{eV}$ between nearest-neighbor electronic densities ${n_{i\sigma} = c^{\dagger}_{i\sigma} c^{\phantom{\dagger}}_{i\sigma}}$ is larger than the half of the on-site Coulomb repulsion, $V_{01}>U/2$.
This suggests that the considered system may have a tendency to form a CDW phase due to the competition between local and non-local Coulomb interactions.
The full form of the long-range Coulomb interaction is presented in the Supplemental Material (SM)~\cite{SM}.

To estimate the phonon properties we utilize the constrained Density Functional Perturbation Theory (cDFPT)~\cite{PhysRevB.92.245108} at hole doping. 
We find that the effective electron-phonon coupling ${\lambda = 2 \int d\omega \, \frac{\alpha^2F(\omega)}{\omega}}$ is dominated by a rather sharp resonance at low phonon frequency, which we can approximate with a local phonon model, i.e ${\alpha^2F(\omega) = N_0 g^2 \delta(\omega - \omega_{\rm ph})}$. Here, $N_0$ is the DOS at the Fermi level, ${\omega_{\rm ph}=8.5\,\text{meV}}$ the phonon energy, and 
${g=34.7\,\text{meV}}$ its coupling strength to the upmost valence band~\cite{SM}.
The strong local coupling of electrons to phonons renders the CDW formation even more favorable.
Indeed, upon integrating out bosonic operators $b^{(\dagger)}$ that correspond to phonon degrees of freedom one gets an effective local frequency-dependent attractive interaction
${U^{\rm ph}_{\omega} = 2g^2\frac{\omega_{\rm ph}}{\omega^2_{\rm ph}-\omega^2}}$~\cite{PhysRevB.52.4806, PhysRevLett.94.026401, PhysRevLett.99.146404}.
This interaction reduces the repulsive on-site Coulomb potential as ${U \to U - U^{\rm ph}_{\omega}}$ and thus enhances the effect of the non-local Coulomb interaction $V_{ij}$.

{\it Method} --- An accurate theoretical investigation of many-body instabilities in InSe monolayer cannot be performed within conventional perturbative methods like the random phase approximation~\cite{pines1966theory, platzman1973waves, vonsovsky1989quantum} or the $GW$ approach~\cite{GW1, GW2, GW3}. 
Local correlation effects that are governed by such a large value of the local Coulomb interaction $U$ should be addressed using at least the dynamical mean-field theory (DMFT)~\cite{RevModPhys.68.13}.
At the same time, spatial collective electronic fluctuations and the long-range Coulomb interaction cannot be taken into account in the framework of DMFT and require to use diagrammatic extensions of this theory~\cite{RevModPhys.90.025003}. 
In this work, we solve the considered many-body problem using the dual triply irreducible
local expansion ($\text{D-TRILEX}$) method~\cite{PhysRevB.100.205115, PhysRevB.103.245123, 2020arXiv201003433S}.
The $\text{D-TRILEX}$ approach is one of the simplest consistent diagrammatic extensions of DMFT that allows one to account for leading collective electronic fluctuations on equal footing without any limitation on the range.
Within this method local correlation effects are treated via the self-energy $\Sigma^{\rm imp}_{\nu}$ and the polarization operator $\Pi^{\varsigma\,\rm imp}_{\omega}$ of an effective local impurity problem of DMFT.
The corresponding impurity problem is solved numerically exactly using the open source CT-HYB solver~\cite{HAFERMANN20131280, PhysRevB.89.235128} based on ALPS libraries~\cite{Bauer_2011}.
The spatial fluctuations are considered in a partially-bosonized form of the renormalized charge (${\varsigma={\rm c}}$) and spin (${\varsigma={\rm s}}$) interactions $W^{\varsigma}_{{\bf q}\omega}$ that enter the diagrammatic part of the self-energy $\overline{\Sigma}_{{\bf k}\nu}$ introduced beyond DMFT. 
Thus, the $\text{D-TRILEX}$ method can be seen as the $GW$ extension of the DMFT that, however, additionally considers exact local three-point vertex corrections in diagrams for the self-energy and the polarization operator~\cite{PhysRevB.100.205115, PhysRevB.103.245123}.
Such vertices are of a crucial importance for a correct description of magnetic, optical and transport properties of the system~\cite{2020arXiv201003433S, Aryasetiawan08, Sponza17, doi:10.1143/JPSJ.75.013703, PhysRevB.80.161105, Katsnelson_2010, PhysRevB.84.085128, Ado_2015, PhysRevLett.117.046601, PhysRevLett.123.036601, PhysRevLett.124.047401, 2020arXiv201009052S}.
The dressed Green's function $G_{{\bf k}\nu}$ of the problem can be found using the standard Dyson equation $G^{-1}_{{\bf k}\nu} = i\nu + \mu - \varepsilon_{\bf k} - \Sigma_{{\bf k}\nu}$ written in momentum ${\bf k}$ and fermionic Matsubara frequency $\nu$ space. 
In this expression $\mu$ is the chemical potential, $\varepsilon_{\bf k}$ is the electronic dispersion that can be obtained as a Fourier transform of the hopping amplitudes $t_{ij}$, and $\Sigma_{{\bf k}\nu} = \Sigma^{\rm imp}_{\nu} + \overline{\Sigma}_{{\bf k}\nu}$ is the total self-energy.
The renormalized interaction $W^{\varsigma}_{{\bf q}\omega}$ can be found via the following Dyson equation ${W^{\varsigma\,-1}_{{\bf q}\omega} = U^{\varsigma\,-1}_{{\bf q}\omega} - \Pi^{\varsigma}_{{\bf q}\omega}}$, where ${U^{\rm c}_{{\bf q}\omega} = U/2 - U^{\rm ph}_{\omega} + V_{\bf q}}$ and ${U^{\rm s}_{{\bf q}\omega} = -U/2}$ are the bare interactions in the charge and spin channels, respectively~\cite{PhysRevB.100.205115, PhysRevB.103.245123}. $\Pi^{\varsigma}_{{\bf q}\omega} = \Pi^{\varsigma\,{\rm imp}}_{\omega} + \overline{\Pi}^{\varsigma}_{{\bf q}\omega}$ is the total polarization operator of the problem, where $\overline{\Pi}^{\varsigma}_{{\bf q}\omega}$ is the diagrammatic contribution introduced in the $\text{D-TRILEX}$ approach~\cite{PhysRevB.100.205115, PhysRevB.103.245123}.
Charge and spin susceptibilities $X^{\varsigma}_{{\bf q}\omega}$ can then be obtained straightforwardly as
${X^{\varsigma\,-1}_{{\bf q}\omega} = \Pi^{\varsigma\,-1}_{{\bf q}\omega} - U^{\varsigma}_{{\bf q}\omega}}$~\cite{2020arXiv201003433S}.

\begin{figure}[t!]
\includegraphics[width=0.95\linewidth]{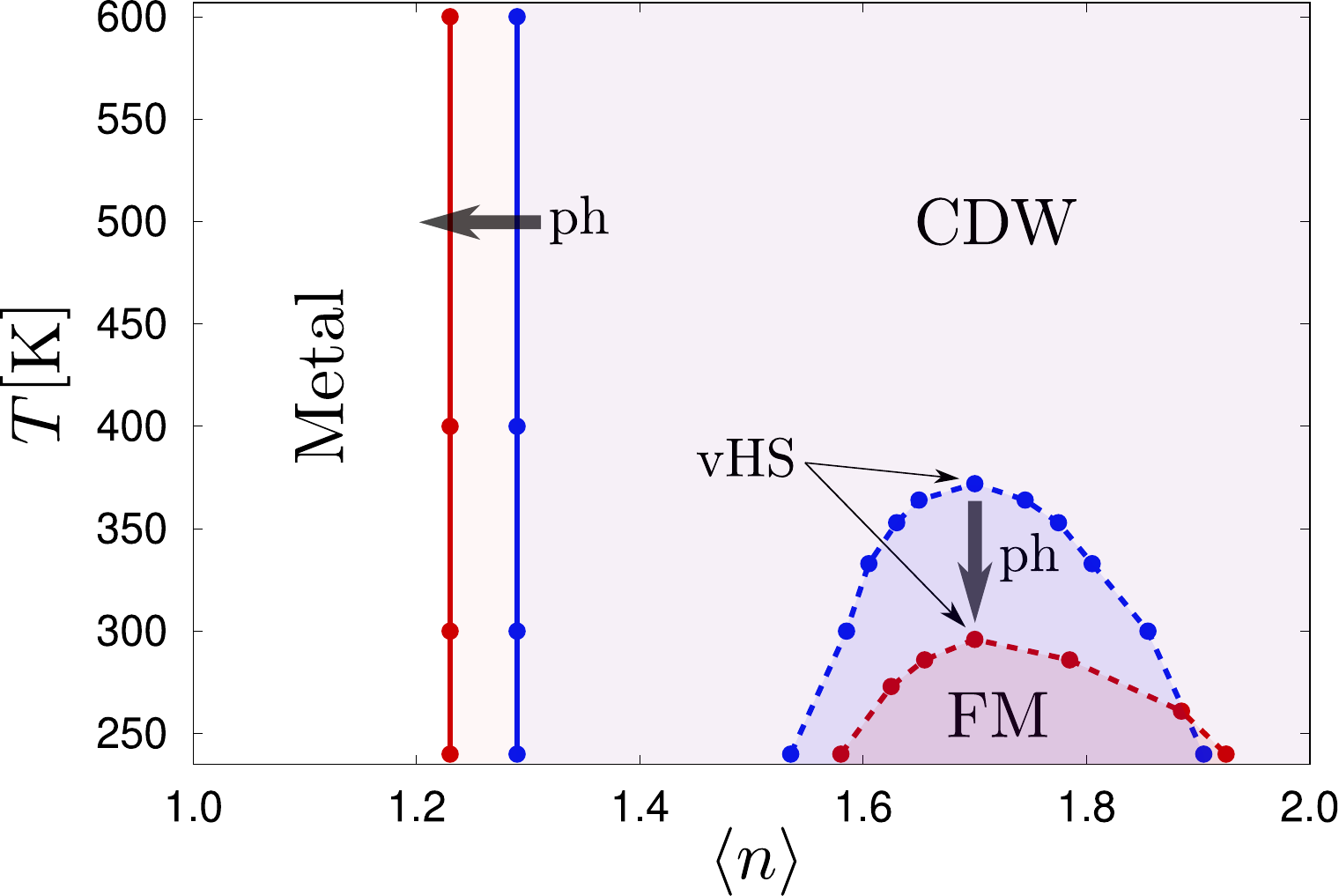}
\caption{\label{fig:phase} Phase diagram for the monolayer InSe as a function of temperature and doping. Solid vertical lines correspond to the CDW phase boundaries, dashed lines depict the FM instabilities. Results are obtained in the presence (red line) and in the absence (blue line) of the electron-phonon coupling. The top of the FM dome corresponds to the filling $\langle n \rangle = 1.70$ at which the vHS appears at the Fermi energy. Black arrows with the label ``ph'' illustrate the effect of phonon degrees of freedom that tend to suppress the FM instability and favor the CDW ordering. }
\end{figure}

{\it Collective electronic instabilities} --- One of the most remarkable features of the InSe monolayer is the presence of a Mexican-hat-like valence band in the electronic dispersion~\cite{doi:10.1063/1.5027023, PhysRevMaterials.3.034004}.
The top of this band exhibits flat regions that lead to a sharp vHS in the DOS.
However, under normal conditions this valence band is fully filled, making the material an indirect semiconductor.
To enhance correlation effects in the system we consider realistic hole dopings with the Fermi level close to the vHS.
Practically, high concentration of holes of the order of $10^{14}$ cm$^{-2}$ can be achieved in 2D materials by means of electrostatic solid- or liquid-electrolyte gating~\cite{Kim} or by surface molecular doping~\cite{Brus}.
First, we solve the many-body problem without taking into account the electron-phonon coupling in order to investigate purely Coulomb correlation effects.
For detecting main instabilities in the system we perform single-shot $\text{D-TRILEX}$ calculations for the charge and spin susceptibilities $X^{\varsigma}_{{\bf q}\omega}$.
This way the diagrammatic part of the polarization operator $\overline{\Pi}^{\varsigma}_{{\bf q}\omega}$ is obtained non-self-consistently in terms of DMFT Green's functions $G^{\rm D}_{{\bf k}\nu}$, which are dressed only by the local self-energies: ${G^{\rm D\,-1}_{{\bf k}\nu} = i\nu + \mu - \varepsilon_{\bf k} - \Sigma^{\rm imp}_{\nu}}$. 
This form of the $\text{D-TRILEX}$ susceptibility resembles the DMFT susceptibility~\cite{PhysRevB.85.115128, Boehnke_2018, PhysRevB.100.125120} with a longitudinal dynamical vertex corrections~\cite{PhysRevB.103.245123}.
In this case, divergences of charge and spin susceptibilities do not affect each other through the self-energy, which allows one to detect instabilities inside broken-symmetry phases.

\begin{figure}[t!]
\includegraphics[width=0.95\linewidth]{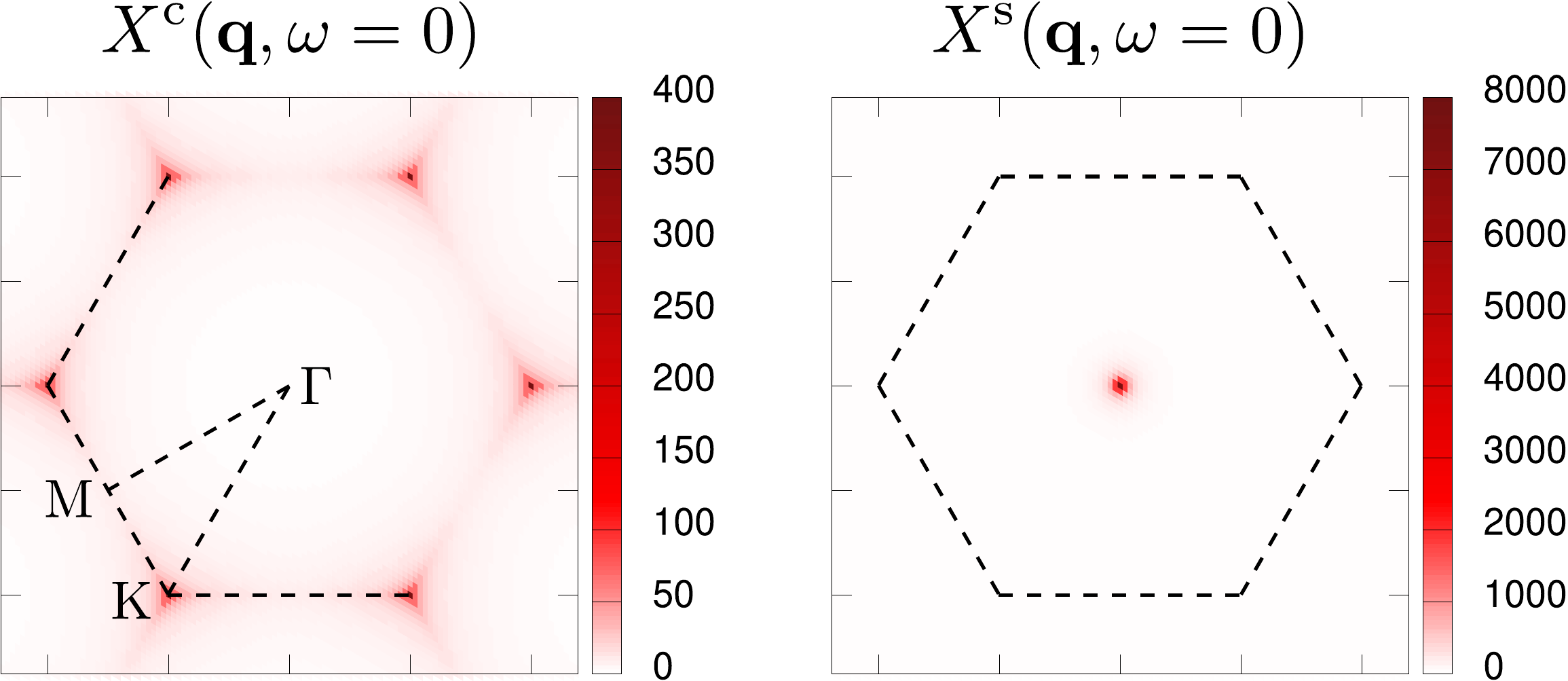}
\caption{\label{fig:X} Momentum resolved charge (left panel) and spin (right panel) susceptibilities ${X^{\varsigma}_{{\bf q},\omega}}$ calculated at zero frequency ${\omega=0}$ in the absence of the electron-phonon coupling. Results are obtained close to the CDW ( ${T=300\,\text{K}}$, ${\langle n \rangle = 1.29}$) and the FM (${T=375\,\text{K}}$, ${\langle n \rangle = 1.70}$) instabilities, respectively. Bragg peaks that appear in the charge susceptibility at the K points of the hexagonal BZ correspond to a commensurate CDW ordering. A single peak at the $\Gamma$ point in the spin susceptibility confirms that the observed instability is FM.}
\end{figure}

Figure \ref{fig:phase} shows the obtained phase diagram for the InSe monolayer, where ${\langle n \rangle}$ is the filling of the considered valence band (${\langle n \rangle}=2$ in the fully filled band that corresponds to the undoped case). 
Phase boundaries indicate points in the temperature $(T)$ vs. doping space, where corresponding susceptibilities diverge.
We find that the charge susceptibility diverges in a broad range of hole dopings ${\langle n \rangle\geq1.29}$, and the corresponding phase boundary is independent of temperature.
The left panel of Fig.~\ref{fig:X} displays the momentum resolved charge susceptibility ${X^{\rm c}_{{\bf q}\omega}}$ obtained at the zero frequency ${\omega=0}$ near the transition point (${T=300\,\text{K}}$, ${\langle n \rangle=1.29}$).
It shows that the corresponding Bragg peaks in the charge susceptibility appear at the K points of the Brillouin zone (BZ), which indicates the formation of a commensurate CDW ordering.
In turn, the spin susceptibility remains finite at the CDW transition point (see SM~\cite{SM}) and diverges only inside the CDW phase.
The corresponding instability has a dome shape as depicted in Fig.~\ref{fig:phase} by a blue dashed line. 
Remarkably, we find that the top of the dome corresponds to the filling ${\langle n \rangle = 1.70}$ at which the vHS appears exactly at the Fermi level. 
The momentum resolved spin susceptibility obtained close to the top of the dome (${T=375\,\text{K}}$, ${\langle n \rangle = 1.70}$) is shown in right panel of Fig.~\ref{fig:X}.
It reveals a sharp Bragg peak at the $\Gamma$ point of the BZ, which indicates the tendency of the system towards FM ordering.

\begin{figure}[t!]
\includegraphics[width=0.85\linewidth]{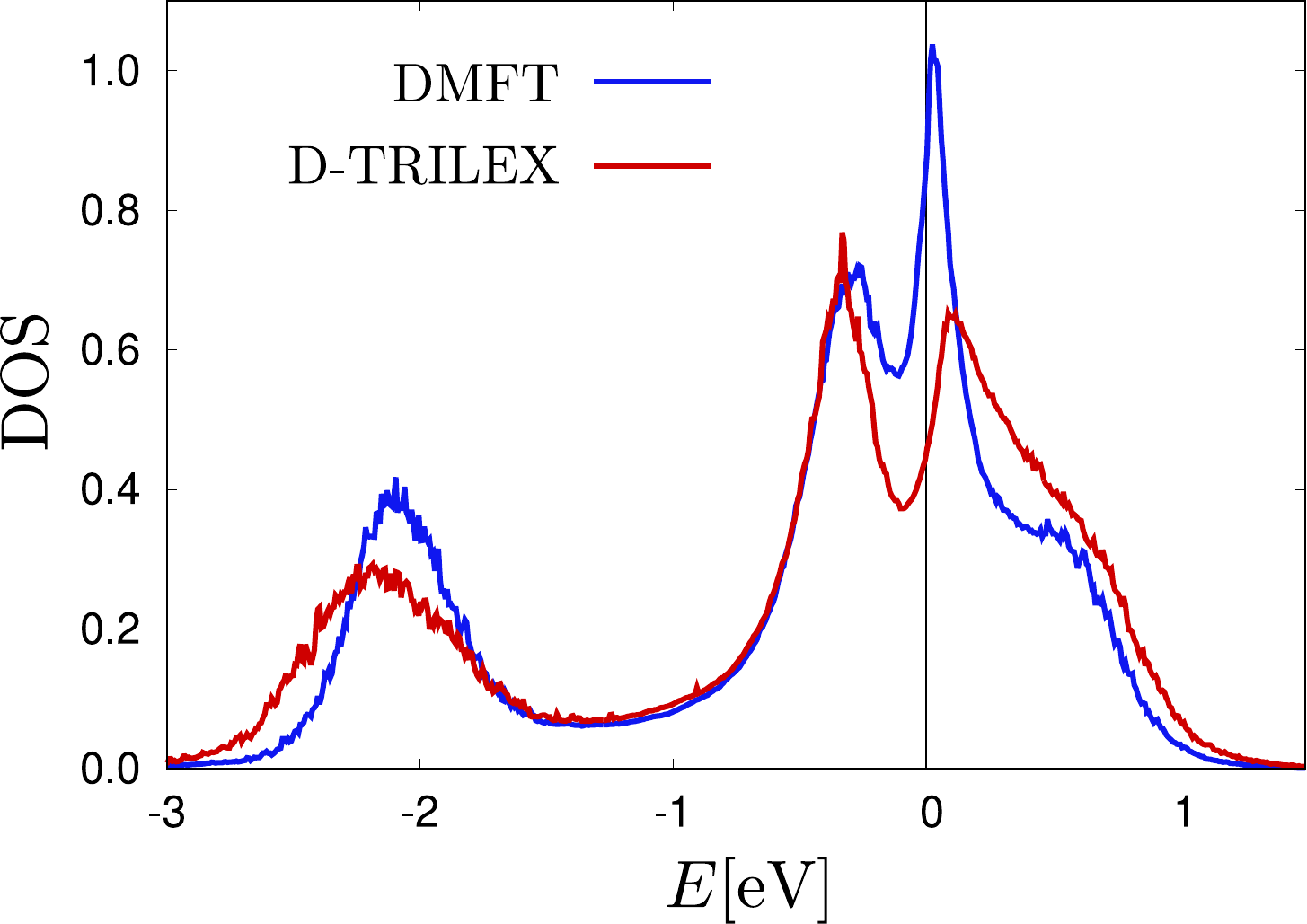}
\caption{\label{fig:DOS} DOS of the InSe monolayer obtained within DMFT (blue line) and D-TRILEX (red line) methods close to a CDW phase boundary at ${T=300\,\text{K}}$ and ${\langle n \rangle = 1.29}$ without taking into account the electron-phonon coupling.}
\end{figure}

The obtained phase diagram illustrates that the commensurate CDW represents the main instability in the InSe monolayer contrary to the DFT prediction~\cite{doi:10.1021/acsanm.8b01476}.
The formation of this insulating phase is associated with strong long-range collective charge fluctuations that are expected to renormalize the electronic dispersion.
Therefore, the development of the CDW ordering should also be reflected in single-particle observables such as the electronic DOS.
In order to account for the feedback effects of long-range collective electronic fluctuations to single-particle quantities we perform self-consistent $\text{D-TRILEX}$ calculations~\cite{PhysRevB.100.205115, PhysRevB.103.245123} and obtain the full lattice Green's function $G_{{\bf k}\nu}$ of the considered many-body problem that takes into account both, local and non-local correlation effects.
Further, we take the local part of the lattice Green's function ${G^{\rm loc}_{\nu} = \frac{1}{N}\sum_{\bf k} G_{{\bf k}\nu}}$ and perform an analytical continuation from Matsubara frequency space $\nu$ to real energies using the stochastic optimization method~\cite{KRIVENKO2019166} to get the DOS.
The obtained result is compared to the one of the DMFT that does not take into account spatial correlation effects.

Figure~\ref{fig:DOS} shows the DOS calculated via both methods close to the CDW phase boundary (${T=300\,\text{K}}$ and ${\langle n \rangle = 1.29}$). 
We find that the DOS obtained within DMFT (blue line) is similar to the one of DFT [Fig.~\ref{fig:bands}(b)] and shows a sharp peak in the vicinity of the Fermi energy, which reflects the presence of the vHS in the electronic spectrum.
However, if one additionally accounts for the effect of spatial correlations via the $\text{D-TRILEX}$ approach, one observes that at the transition point this peak turns into a pseudogap (red line).
In this particular case, the pseudogap appears due to strong collective charge fluctuations that lead to an almost diverged charge susceptibility ${X^{\rm c}_{{\bf q}\omega}}$ at the ${\bf q}=\text{K}$ point of the BZ close to a phase transition.
As a consequence, the renormalized interaction $W^{\rm c}_{{\bf q}\omega}$ that enters the self-energy also becomes nearly divergent, which causes the formation of a pseudogap according to the tendency towards electron-hole pairing.
This mechanism is similar to the formation of the excitonic insulator state~\cite{keldysh, kozlov, HALPERIN1968115} with the only difference that in our case electrons and holes belong to the same band and that excitons in the condensate have non-zero momentum corresponding to the CDW wave vector. 
If the tendency towards electron-hole pairing results in long-range order, one gets a true gap whereas strong short-range order without long-range order leads to a pseudogap in the electronic spectrum~\cite{Irkhin_1991, PhysRevLett.120.216401}.
This result illustrates the importance of a self-consistent consideration of long-range collective electronic fluctuations for capturing the formation of the insulating CDW phase in monolayer InSe.

In order to investigate the effect of phonon degrees of freedom on the observed instabilities we repeat the same calculation in the presence of the electron-phonon coupling.
We find that in this case the CDW phase boundary is shifted to smaller values of the filling $\langle n \rangle = 1.23$. 
At the same time, the FM instability is pushed down to lower temperatures, but the top of the FM dome remains at the vHS filling $\langle n \rangle = 1.70$ as in the absence of phonons.
This result is consistent with the presented argument above that the electron-phonon coupling effectively reduces the on-site Coulomb potential, which consequently decreases the critical temperature for the magnetic instability. 
The observed shift of the CDW phase boundary can also be explained by the same argument.
Indeed, the local Coulomb repulsion favors the single occupation of lattice sites.
On the contrary, the non-local Coulomb interaction promotes CDW ordering, which upon reducing the local Coulomb interaction becomes energetically preferable.
Remarkably, we find that the position of the Bragg peaks in the charge and spin susceptibilities calculated close to corresponding phase boundaries remain unchanged~\cite{SM}.
In addition, we do not find any additional Bragg peak in the charge susceptibility at twice the Fermi wave vector $2k_{F}$, which would indicate a charge instability due to the electron-phonon mechanism.
This fact suggests that many-body effects in the monolayer InSe are driven by strong Coulomb correlations rather than by the electron-phonon mechanism as suggested previously~\cite{PhysRevB.103.035411}.

{\it Conclusions} ---
We have systematically studied many-body effects in the hole-doped InSe monolayer.
We have found that this material displays coexisting instabilities that are mainly driven by nonlocal Coulomb correlations.
The commensurate CDW ordering represents the main instability in the system and is revealed in a broad range of doping levels and temperatures.
This ordering is accompanied by the formation of an insulating phase that is confirmed by the appearance of a pseudogap in the DOS close to the transition point, which illustrates the importance of considering spatial electronic fluctuations.
We also observed the tendency to a FM ordering that manifests itself only inside the CDW phase and is related to a vHS in the electronic spectrum.
The inclusion of the electron-phonon coupling results in a shift of the CDW and FM ordering phases on the phase diagram, which can be explained by an effective reduction of the local Coulomb interaction.
However, the qualitative physical picture does not change in the presence of phonon degrees of freedom.
Our results suggest that monolayer InSe can serve as an attractive playground for investigation of coexisting many-body correlation effects and, in particular, of 2D magnetism, although in a bulk phase this material is non-magnetic.\\

\begin{acknowledgments}
We thank Andy J. Millis for fruitful discussions and Jan Berges for sharing his cDFPT-related changes to {\sc Quantum Espresso} with us.
The work of E.A.S. was supported by the European Union’s Horizon 2020 Research and Innovation programme under the Marie Sk\l{}odowska Curie grant agreement No.~839551 - $\text{2DMAGICS}$. The work of M.I.K., A.N.R., and A.I.L. was supported by European Research Council via Synergy Grant 854843 - FASTCORR. V.H. and A.I.L. acknowledge the support by the Cluster of Excellence ``Advanced Imaging of Matter'' of the Deutsche Forschungsgemeinschaft (DFG) - EXC 2056 - Project No.~ID390715994. E.A.S., V.H., and A.I.L. also acknowledge the support by North-German Supercomputing Alliance (HLRN) under the Project No.~hhp00042.
\end{acknowledgments}

\onecolumngrid
\begin{center}
\begin{large}
\textbf{~\\~\\~\\
Supplemental Material\\[0.5cm]
Coexisting charge density wave and ferromagnetic instabilities in monolayer InSe}\\[1cm]
\end{large}
\end{center}
\twocolumngrid

\begin{figure}[t!]
\includegraphics[width=0.95\linewidth]{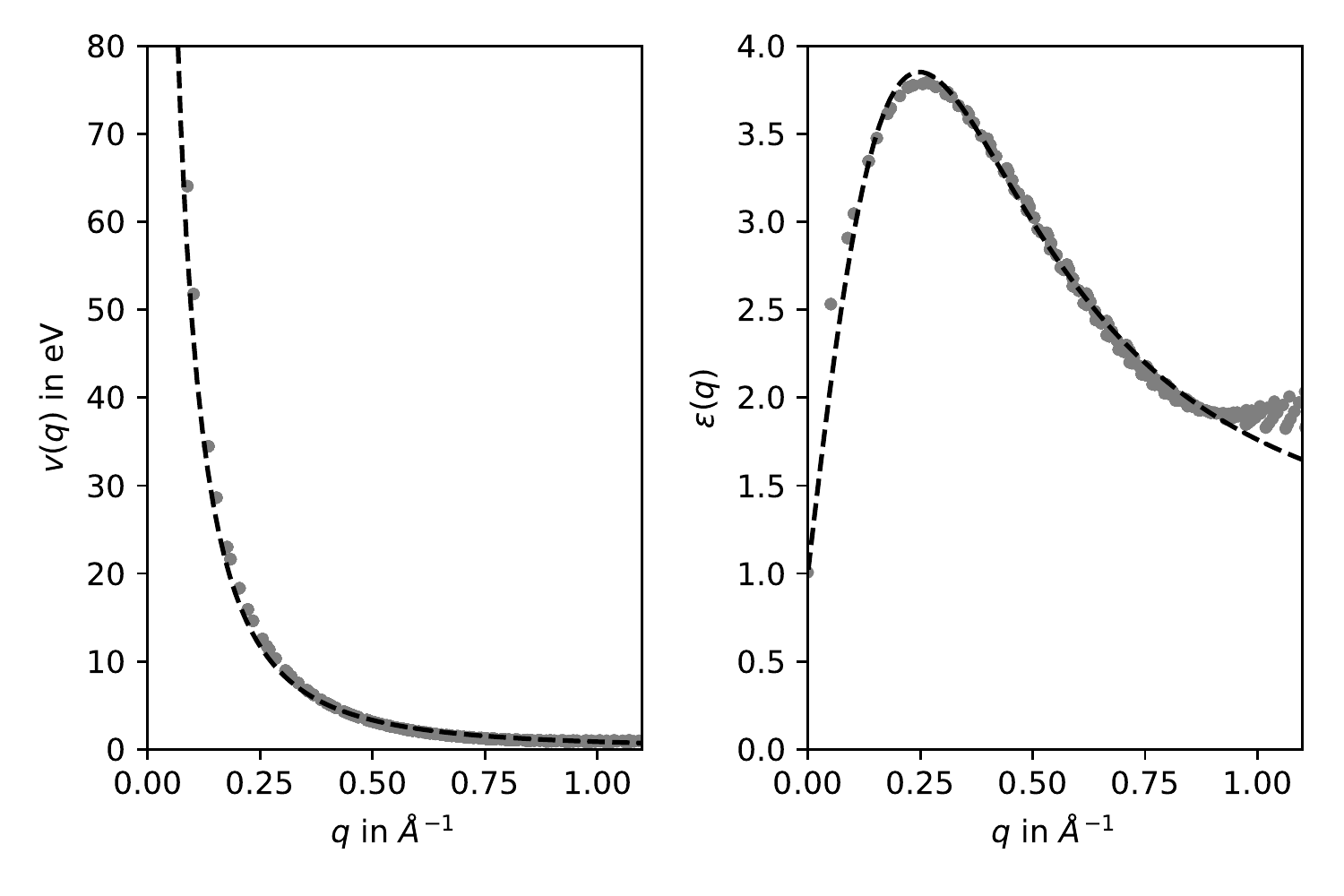}
\caption{\label{fig:CM_SM} Bare Coulomb interaction (left) and cRPA dielectric function (right) within the Wannier basis. Markers represent ab initio data and dashed lines the corresponding fits.}
\end{figure}

\section{Details of first-principles calculations}
The band structure presented in Fig.~1(b) was calculated within density functional theory utilizing the projected augmented wave (PAW) formalism~\cite{paw1, paw2} as implemented in the \emph{Vienna ab initio simulation package} ({\sc vasp})~\cite{Kresse1, Kresse2}. The exchange-correlation effects were considered using the generalized gradient approximation (GGA)~\cite{gga}. A 500 eV energy cut-off for the plane-waves and a convergence threshold of 10$^{-7}$ eV were used in the calculations. The Brillouin zone was sampled by a (36$\times$36) ${\bf  k}$-point mesh. We adopted fully relaxed atomic structure with a lattice constant of 3.94~\AA. A $\sim$30~\AA-thick vacuum layer was added in the direction perpendicular to the 2D plane in order to avoid spurious interactions between supercell images.
The Wannier functions and the tight-binding Hamiltonian were calculated within the scheme of maximal localization~\cite{mlwf1,mlwf2} using the {\sc wannier90} package~\cite{wannier90}. 

\begin{figure}[t!]
\includegraphics[width=0.95\linewidth]{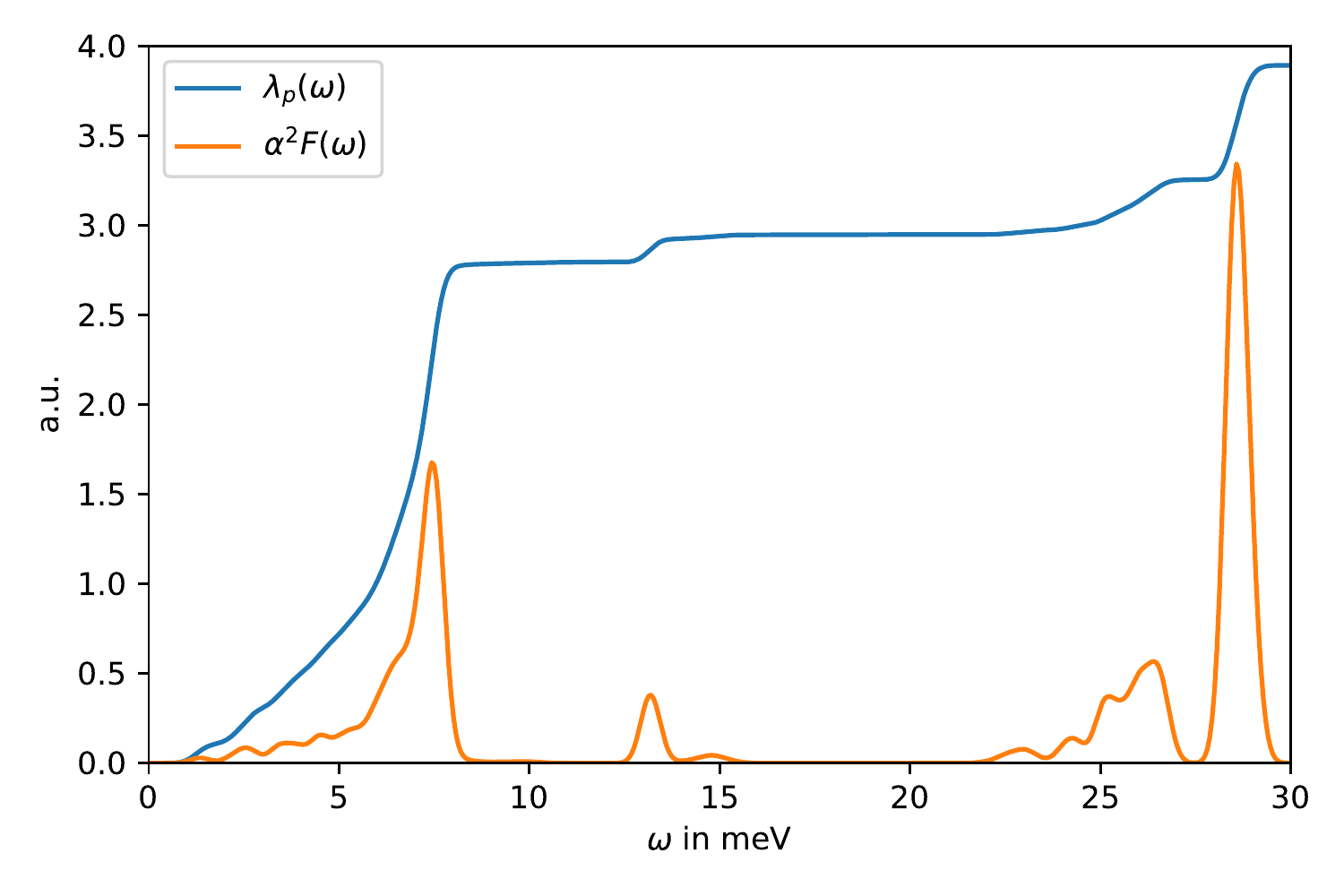}
\caption{\label{fig:Phonon_SM} cDFPT Eliashberg function together with the partial effective electron-phonon coupling $\lambda_p(\omega) = 2 \int_0^\omega d\omega' \, \frac{\alpha^2F(\omega')}{\omega'}$.}
\end{figure}

The Coulomb interaction was evaluated using the maximally localized Wannier functions within the constrained random phase approximation (cRPA)~\cite{CoulombU, KaltakcRPA} as ${U_{ij}=\langle w_i w_j |U|w_j w_i\rangle}$, where $U$ is the partially screened Coulomb interaction defined by ${U=v+v \Pi U}$ with $v$ being the bare Coulomb interaction, $\Pi$ the cRPA polarization, and $w_i$ is the Wannier function at the lattice site $i$. The polarization operator $\Pi$ describes screening from all electronic states except those given by the tight-binding Hamiltonian obtained in the Wannier basis. For these calculations, we used a recent cRPA implementation by Kaltak within {\sc vasp}~\cite{KaltakcRPA}. To converge the cRPA polarization with respect to the number of empty states with used in total $64$ bands. To derive a light-weighted Coulomb model for arbitrary $q$ grids, we fitted the cRPA Coulomb interaction in momentum space according to:
\begin{align}
    U(q) = \frac{v(q)}{\varepsilon(q)} \label{eqn:U_SM}
\end{align}
with the bare Coulomb interaction of a monolayer
\begin{align}
    v(q) = \frac{h}{2\pi} \int_{-\pi/h}^{+\pi/h} \frac{4\pi e^2}{Vq^2}
         = \frac{4e^2}{A} \frac{\operatorname{arctan}\left( \frac{\pi}{qh} \right)}{q}
\end{align}
and the dielectric function
\begin{align}
    \varepsilon(q) = \varepsilon_0(q)
                     \frac{\varepsilon_0(q) + 1 - (\varepsilon_0(q) - 1) e^{-qd}}{\varepsilon_0(q) + 1 + (\varepsilon_0(q) - 1) e^{-qd}}
\end{align}
with
\begin{align}
    \varepsilon_0(q) = \frac{a + q^2}{\frac{a \sin(qc)}{bqc} + q^2}.
\end{align}
Here, $e$, $A$, and $h$ are the elementary electron charge, the InSe unit cell area and its effective height, respectively, and $a$, $b$, $c$, and $d$ are fitting parameter. In Fig.~\ref{fig:CM_SM} we show the corresponding ab initio values together with their fits using:
\begin{align*}
    h = 15.35\, \AA \qquad & a = 1.07\, \AA^{-2} \\
                  & b = 5.79 \\
                  & c = 0.14\, \AA \\
                  & d = 8.09\, \AA
\end{align*}
The real-space $U_{ij}$ are calculated via a coventional Fourier transform of $U(q)$ from Eq.~(\ref{eqn:U_SM}) and using these fits.

\begin{figure}[t!]
\includegraphics[width=0.45\linewidth]{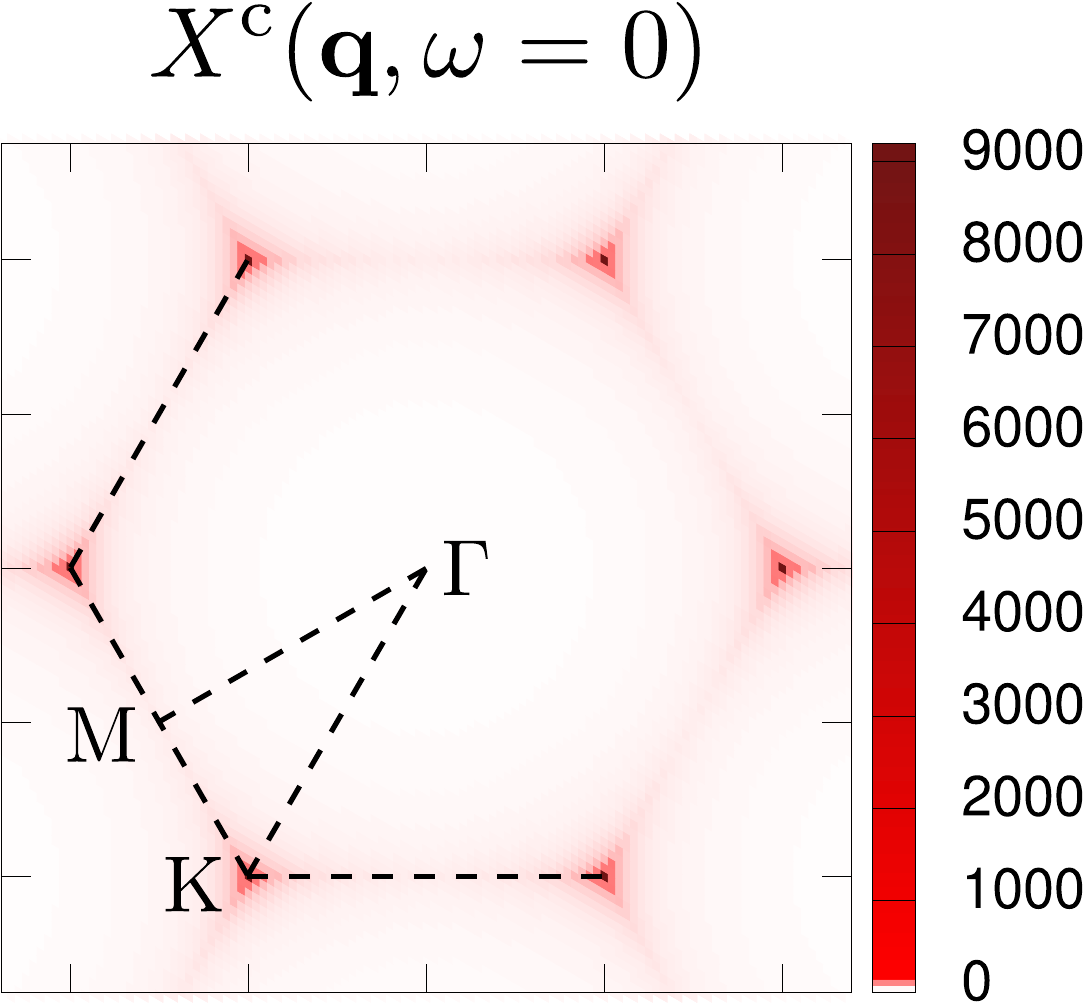}
\caption{\label{fig:Xc_SM} Momentum resolved charge susceptibility ${X^{\rm c}({\bf q}, \omega=0)}$ calculated taking into account the electron-phonon coupling. Result is obtained close to the CDW transition point ${T=300\,\text{K}}$ and ${\langle n \rangle = 1.23}$.  
}
\end{figure}

The phonon properties are calculated within {\sc Quantum Espresso}~\cite{Giannozzi_2017} using norm-conserving pseudopotentials, the local density approximation, an energy cutoff of $80\,$Ryd, and a finite hole doping of $0.04$ holes per unit cell. For the intial constrained Density Functional Perturbation Theory (cDFPT)~\cite{PhysRevB.92.245108} calculation we use (16$\times$16) ${\bf  k}$ and (8$\times$8) ${\bf  q}$-point meshes and exclude the highest (hole doped) valence band within the evaluation of the Sternheimer equation. Afterwards we use the {\sc EPW} code~\cite{PONCE2016116} to extrapolate the cDFPT results to (32$\times$32) ${\bf  k}$- and  ${\bf  q}-$point meshes using a Wannier interpolation based on a single Wannier projection for the highest valence band (in the very same way as we constructed them in {\sc vasp} for the cRPA calculations). This allows us to accurately calculate the cDFPT Eliashberg function $\alpha^2F(\omega)$ as shown in Fig.~\ref{fig:Phonon_SM}. From this we calculate the effective electron-phonon coupling ${\lambda = 2 \int d\omega \, \frac{\alpha^2F(\omega)}{\omega}}$ which we finally use to fit our local phonon model via ${\alpha^2F(\omega) \approx N_0 g^2 \delta(\omega - \omega_{\rm ph})}$ with ${N_0 = 3.88\,}$states/spin/eV/unit cell.

\section{Susceptibility}

In this section we show the results for the charge and spin susceptibilities obtained within single-shot D-TRILEX calculations~\cite{PhysRevB.100.205115, PhysRevB.103.245123} in the presence of the electron-phonon coupling.
As discussed in the main text, taking into account phonon degrees of freedom shifts the charge density wave (CDW) phase boundary to a smaller value of the filling $\langle n \rangle = 1.23$ compared to the one $\langle n \rangle = 1.29$ obtained in the absence of the electron-phonon coupling.
The charge susceptibility calculated close to the CDW phase transition point ${T=300\,\text{K}}$ and ${\langle n \rangle = 1.23}$ is presented in Fig.~\ref{fig:Xc_SM}. 
The structure of the susceptibility in momentum space consists of six delta-function-like Bragg peaks that appear at K points of the Brillouin zone (BZ). 
This fact illustrates that the developed charge ordering corresponds to a commensurate CDW.
Remarkably, the form of the charge susceptibility calculated with the electron-phonon coupling is similar to the one obtained neglecting phonon degrees of freedom (see main text) and does not show any additional features, like Bragg peaks at twice the Fermi wavevector $2k_{F}$, that could be associated with the effect of phonons. 
This result suggests that the CDW instability in the monolayer InSe is driven by strong electronic Coulomb correlations.

\begin{figure}[t!]
\includegraphics[width=0.95\linewidth]{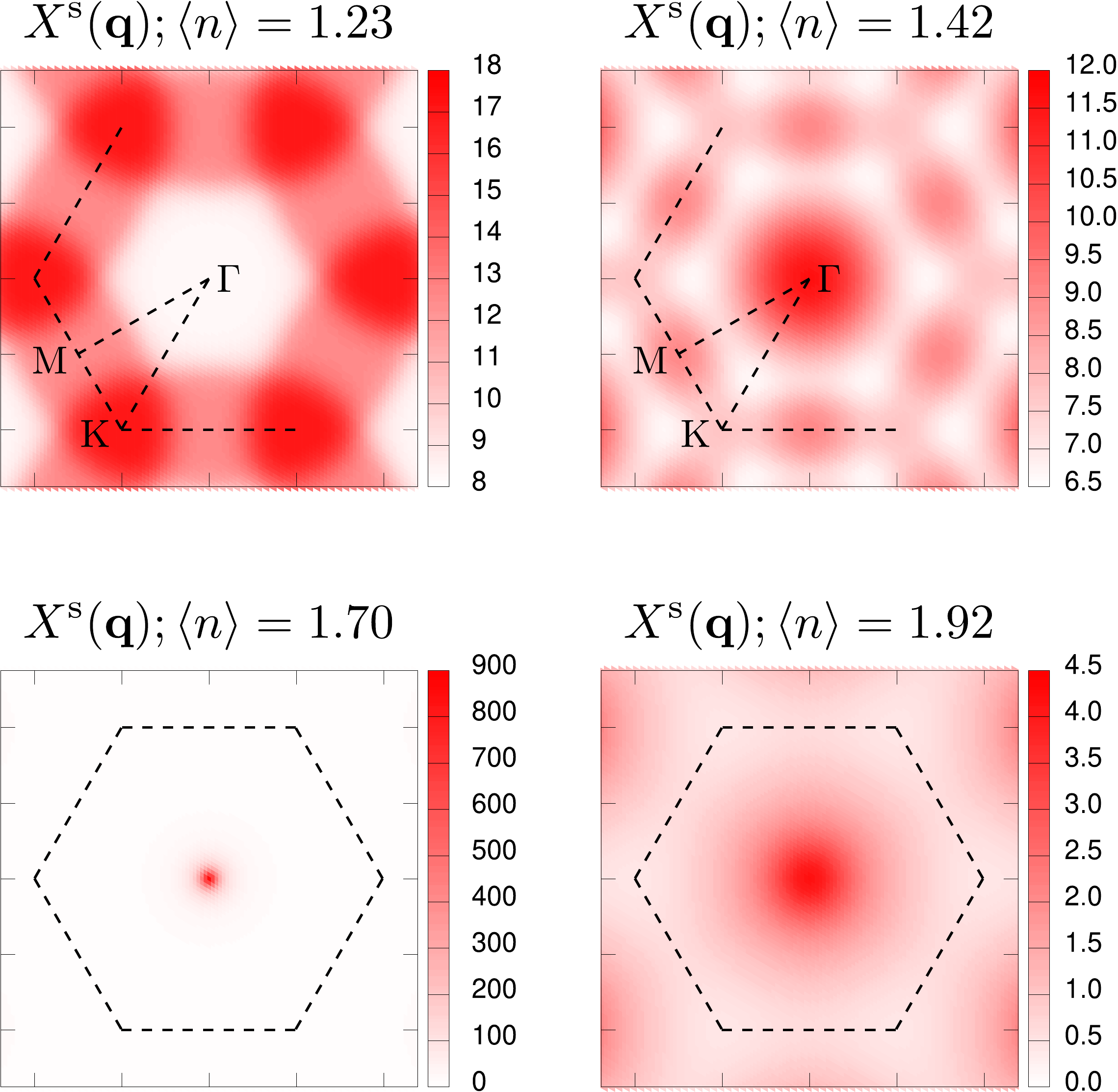}
\caption{\label{fig:Xs_SM} Momentum resolved spin susceptibility ${X^{\rm s}({\bf q})}$ calculated at zero frequency ${\omega=0}$ taking into account the electron-phonon coupling. Results are obtained at ${T=300\,\text{K}}$ for different values of the filling ${\langle n \rangle = 1.23}$ (top left panel), ${\langle n \rangle = 1.42}$ (top right panel), ${\langle n \rangle = 1.70}$ (bottom left panel), and ${\langle n \rangle = 1.92}$ (bottom right panel). }
\end{figure}

At the CDW transition point the spin susceptibility stays finite and has an antiferromagnetic (AFM) behavior that manifests itself in the largest value of the susceptibility at K point of the BZ (see top left panel of Fig.~\ref{fig:Xs_SM}).
At a larger filling ${\langle n \rangle = 1.42}$ the AFM form of the spin susceptibility changes to a more complex structure with the largest value at $\Gamma$ point of the BZ, which indicates the ferromagnetic instability (FM), and less pronounced intensities at M points (see top right panel of Fig.~\ref{fig:Xs_SM}).
Increasing the filling to ${\langle n \rangle = 1.70}$ the van Hove singularity (vHS) in the electronic spectral function appears at the Fermi energy, which strongly enhances collective electronic fluctuations. 
Bottom left panel of Fig.~\ref{fig:Xs_SM} shows that close to the magnetic instability the spin susceptibility becomes purely FM, which is indicated by the single delta-function-like Bragg peak that at the $\Gamma$ point of the BZ.
At even larger filling of the band ${\langle n \rangle = 1.92}$ the spin susceptibility remains FM, but the value of the susceptibility at $\Gamma$ point decreases compared to the case of the vHS filling, which confirms the dome-like structure of the FM instability. 

\begin{figure}[t!]
\includegraphics[width=0.85\linewidth]{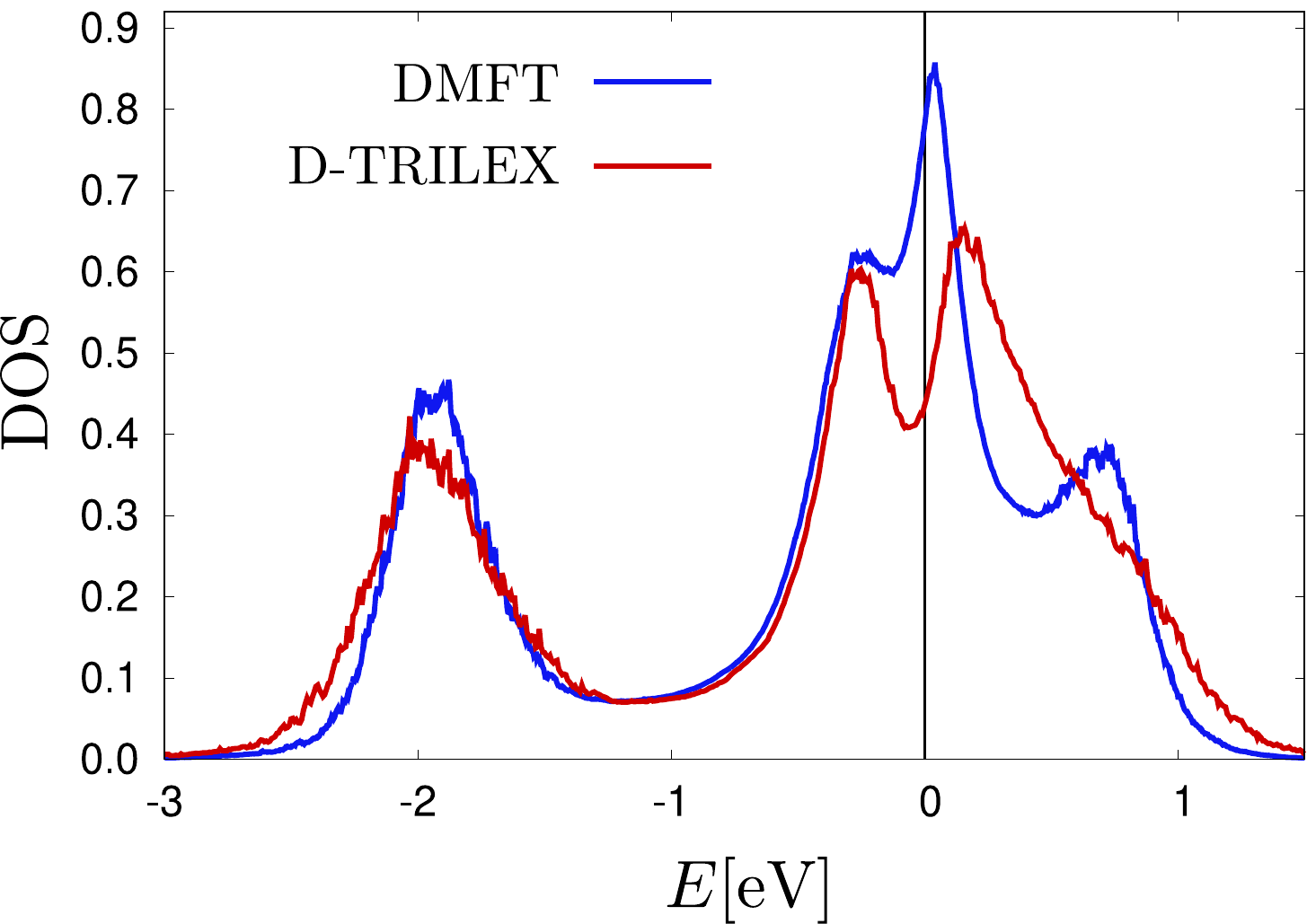}
\caption{\label{fig:DOS_SM} DOS of the monolayer InSe obtained within DMFT (blue line) and D-TRILEX (red line) methods close to a CDW phase boundary at ${T=300\,\text{K}}$ and ${\langle n \rangle = 1.23}$ taking into account the electron-phonon coupling.}
\end{figure}

\section{Local density of states}

In this section we show the density of states (DOS) obtained via the self-consistent D-TRILEX calculation at the CDW phase transition point ${T=300\,\text{K}}$ and ${\langle n \rangle = 1.23}$ in the presence of the electron-phonon coupling.
The corresponding result is shown in Fig.~\ref{fig:DOS_SM} (red line), and is compared to the one of the dynamical mean-field theory (DMFT) (blue line).
Note that DMFT does not take into account spatial correlations, but considers the effect of the local electron-phonon coupling via the renormalized on-site Coulomb potential $U^{\ast} = U - U^{\rm ph}_{\omega}$.  
We find that the DOS predicted by DMFT has a peak close to the Fermi energy, which corresponds to the vHS of the Mexican-hat-like band.
However, at the CDW transition point collective charge fluctuations are strong.
Taking them into account via D-TRILEX method turns this peak into a pseudogap, which confirms the formation of the insulating ordered state in the system.
Remarkably, we find that considering phonon degrees of freedom does not change a qualitative structure of the DOS obtained close to the CDW phase transition point. 

\bibliography{Ref}

\end{document}